\documentclass[twocolumn,runningheads,natbib]{svjour2}
\smartqed  
\usepackage{graphicx}
\journalname{Astrophysics and Space Science}
\begin{document}

\title{On the peculiarities in the rotational frequency evolution of isolated neutron stars
\thanks{ This work has been supported by the Russian Foundation for Basic Research 
(grant No 04-02-17555), Russian Academy of Sciences (program "Evolution 
of Stars and Galaxies"), and by the Russian Science Support Foundation. Also 
the authors are thankful to the anonymous referee for
valuable 
comments.}}

\author{Anton Biryukov         \and
        Gregory Beskin         \and
        Sergey Karpov 
}

\institute{A. Biryukov \at
	      SAI MSU, 13, Universitetsky pr., Moscow, 119992, Russia \\
              \email{eman@sai.msu.ru}            
}

\date{Received: date / Accepted: date}
\maketitle
\begin{abstract}

  The measurements of pulsar frequency second derivatives have shown that they
  are $10^2-10^6$ times larger than expected for standard pulsar spin-down law,
  and are even negative for about half of pulsars.
  We explain these paradoxical results on the basis   of
  the statistical analysis of the rotational parameters $\nu$, $\dot \nu$
  and $\ddot \nu$ of the subset of 295 pulsars taken mostly from the ATNF
  database. We have found a strong correlation between $\ddot \nu$ and
  $\dot \nu$ for both $\ddot\nu > 0$
  and $\ddot\nu < 0$
  , as well as between $\nu$ and $\dot\nu$
  . We interpret these dependencies as evolutionary ones
  due to
  $\dot\nu$ being nearly proportional to the pulsars' age.
  The derived statistical relations as well as ``anomalous'' values of
  $\ddot\nu$ are well described by assuming the long-time variations of the
  spin-down rate. The pulsar frequency evolution, therefore, consists of
  secular change of $\nu_{ev}(t)$, $\dot\nu_{ev}(t)$ and $\ddot\nu_{ev}(t)$
  according to the power law with $n\approx5$, the irregularities, observed
  within a timespan as a timing noise, and the 
  variations on the
  timescale larger than that timespan -- several tens of years.

\end{abstract}

\section{Introduction}
\label{intro}
The spin-down of radio pulsars is due to the conversion of their rotation 
energy into emission. According to the "classical" approach, their rotational 
frequencies $\nu$ evolve obeying the spin-down law $\dot\nu=-K\nu^n$, where 
$K$ is a positive constant that depends on the magnetic dipole moment and the 
moment of inertia of the neutron star, and $n$ is the braking index. The 
latter can be determined observationally from measurements of $\nu$, 
$\dot \nu$ and $\ddot \nu$ as $n={\nu \ddot \nu}/{\dot \nu^2}$. For a simple 
vacuum dipole model of pulsar magnetosphere $n=3$; the pulsar wind decreases 
this value to $n=1$; for multipole magnetic field $n \ge 5$ \citep{man77}. At 
the same time the measurements of pulsar frequency second derivatives 
$\ddot \nu$ have shown that their values are much larger than expected for 
standard spin-down law and are even negative for about half of all pulsars. 
The corresponding braking indices range from $-10^6$ to $10^6$ 
\citep{d'a93, chu03, hob04}.

It was found that the significant correlations between $|\ddot\nu|$ 
($|\ddot P|$) and $\dot\nu$ ($\dot P$) demonstrate the increase of the 
absolute values of the $\nu$ and $P$ second derivatives for younger (with 
faster slow-down) pulsars \citep{cor85, arz94, lyn99}. The anomalously high 
and negative values of $\ddot\nu$ and $n$ may be interpreted as a result of 
low-frequency terms of the ``timing noise'' -- a complex variations of 
pulsars rotational phase
 within a timespan
\citep{d'a93}.

It is clear that the timespan of observations is by no means intrinsic to the
pulsar physics.  Indeed, the variations of rotational parameters may take 
place on larger timescales as well. However, the timescale of observations 
naturally divides it into two separate classes of manifestations -- the 
residuals in respect to the best fit for the timing solution (the ``timing 
noise'') and the systematic shift of the best fit coefficients (i.e. in the 
measured values of $\nu$, $\dot\nu$, $\ddot\nu$) relative to some mean or 
expected value from the model.  The latter effect may be called the ``large 
timescale timing noise''.

Up to date we know nearly 200 pulsars for which the timespan of observations 
is greater than 20 years, and the values of their $\ddot \nu$ still turn out 
to be anomalously large \citep{hob04}.

For example, for the {\it PSR B1706-16} pulsar, variations of $\ddot \nu$ 
with an amplitude of $10^{-24}$ s$^{-3}$ have been detected on a several 
years timescale (see Fig.7 in \citep{hob04}), with the value of $\ddot\nu$ 
depending on the time interval selected. However, the fit over the entire 25 
year timespan gives a value of $\ddot\nu=3.8\cdot10^{-25}$ s$^{-3}$ with a 
few percent accuracy (which leads to a braking index $\approx 2.7\cdot10^3$). 

In the current work we provide observational evidence of the nonmonotonic
evolution of pulsars on timescales larger than the typical contemporary 
timespan of observations (tens of years), using the statistical analysis of
measured $\nu$, $\dot\nu$ and $\ddot \nu$. We estimate the main parameters of 
such long timescale variations and discuss their possible relation to the 
low-frequency terms of the timing noise. We have also derived the parameters 
of pulsar secular spin-down.

\section{Statistical analysis of the ensemble of pulsars}
\label{statan}
Our statistical analysis is based on the assumption that numerous 
measurements of the pulsar frequency second derivatives reflect their 
evolution on the timescale larger than the duration of observations, and uses 
the parameters of 295 pulsars.

\begin{figure}[t]
{\centering \resizebox*{1\columnwidth}{!}{\includegraphics[angle=270]
{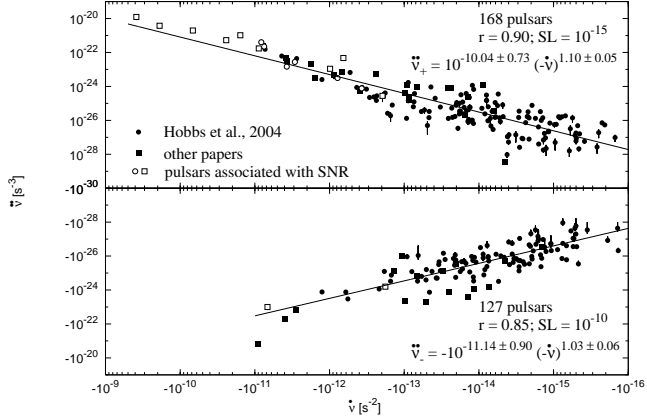}} \par}
\caption{The $\ddot \nu - \dot \nu$ diagram for 295 pulsars. The figure 
shows the pulsars from the work \citep{hob04} as circles, and the objects 
measured by other groups as squares. Open symbols represent the pulsars 
associated with supernova remnants, and therefore -- relatively young ones.
Analytical fits for both positive and negative branches are shown as solid 
lines. Measurment errors shown as an error bars.}
\label{fig1}
\end{figure}

From 389 objects of ATNF catalogue \citep{man05} with known $\ddot \nu$ we 
have compiled a list of "ordinary" radio pulsars with $P>20$ ms and 
$\dot P >10^{-17}$ s s$^{-1}$, excluding recycled, anomalous and binary 
pulsars, and with relative accuracy of second derivative measurements better 
than 75\%. It has been appended with 26 pulsars from other sources 
\citep{d'a93, chu03}. The parameters of all pulsars have been plotted on the 
$\ddot \nu - \dot \nu$ diagram (Fig. \ref{fig1}). 

The basic result of the statistical analysis of this data is a significant
correlation of $\ddot \nu$ and $\dot \nu$, both for 168 objects with $\ddot
\nu > 0$ (correlation coefficient $r\approx0.90$) and for 127 objects with
$\ddot \nu < 0$ ($r\approx0.85$). 
Both groups follow nearly linear laws,
however they seem to be not exactly symmetric relative to $\ddot 
\nu=0$.
We divided both branches into 6 intervals of $\dot \nu$, computed mean values
and its standard deviations
of $\ddot\nu_{\pm}$ in each and rejected the hypothesis of their symmetry 
with a $0.04$ significance level. The absolute values of $\ddot\nu_{+}$ are
systematically larger than corresponding $\ddot\nu_{-}$ (the difference is
positive in 5 intervals of 6). Also, the difference of analytical fits to branches is
positive over $-10^{-11} \div -10^{-15}$ s$^{-2}$ interval of $\dot\nu$.
These are the arguments in favour of a small positive asymmetry of branches.

We found an obvious correlation of $\dot \nu$ with the characteristic age
$\tau_{ch} = -\frac12 \frac{\nu}{\dot \nu}$ ($r=0.96$, 
$\dot\nu\sim\tau_{ch}^{-1.16\pm0.02}$). These parameters are nearly 
proportional, which leads to a significant correlation of $\tau_{ch}$ both 
with $\ddot\nu$ ($r=0.85$ for the positive branch and $r=0.75$ for the 
negative one) and with $n$ ($r=0.75$ and $r=0.76$ correspondingly).

The correlations found are fully consistent with the results published in 
\cite{cor85, arz94, lyn99}, as well as in \cite{ura06}. However, the 
branches with $\ddot \nu >0$ and $\ddot \nu < 0$ in those works were not 
analyzed separately from each other (not as $|\ddot \nu|$).

Young pulsars confidently associated with supernova remnants are 
systematically shifted to the left in  Fig. \ref{fig1} (open symbols). The 
order of their physical ages roughly corresponds to that of their 
characteristic ages. This means that any dependence on $\dot \nu$ or 
$\tau_{ch}$ reflects the dependence on pulsar age.

The $\ddot\nu - \dot \nu$ diagram (Fig. \ref{fig1}) may be 
interpreted as an evolutionary one. In other words, each pulsar during its 
evolution moves along the branches of this diagram while increasing the value 
of its $\dot \nu$ (which corresponds to the increase of its characteristic 
age). However, there is an obvious contradiction: for the negative branch, 
$\dot \nu$, being negative, may only decrease with time (since $\ddot \nu$ is 
formally the derivative of $\dot \nu$), and the motion along the negative 
branch may only be backward! This contradiction is easily solved by assuming 
non-monotonic behaviour of $\ddot\nu(t)$, which has an irregular component 
($\delta \ddot \nu$) along with the monotonic one ($\ddot \nu_{ev}$), where 
the subscript ``ev'' marks the evolutionary value. 
In such 
interpretation the value of $\ddot \nu_{ev}$ must be positive, which will 
lead to a positive asymmetry of the branches. Moreover, the evolutionary 
increasing of $\dot \nu$ implies a positive evolutionary value of $\ddot \nu$.

The characteristic timescale $T$ of such variations must be much shorter than 
the pulsar life time and at the same time much larger than the timescale of 
the observations. As it evolves, a pulsar repeatedly changes sign of 
$\ddot \nu$, in a spiral-like motion from branch to branch, and  spends 
roughly half its lifetime on each one. The asymmetry of the branches reflects 
the positive sign of $\ddot \nu_{ev}(t)$, and therefore, secular increase of 
$\dot\nu_{ev}(t)$ (i.e. all pulsars in their secular evolution move to the 
right on the $\ddot\nu - \dot \nu$ diagram). Systematic decrease of branches 
separation reflects the decrease of the variations amplitude and/or the 
increase of its characteristic timescale. 

Any well known non-monotonic variations of $\dot\nu(t)$, like glitches, 
microglitches, timing noise or precession, will manifest themselves in a 
similar way on the $\ddot\nu - \dot \nu$ diagram and lead to extremely high 
values of $\ddot \nu$ \citep{she96, sta00}. However, their characteristic
timescales vary from weeks to years, and they are detected immediately. But 
here the variations on much larger timescales are discussed, and their study 
is possible only statistically, assuming the ergodic behaviour of the ensemble
of pulsars.

\section{Non-monotonic variations of pulsar spin-down rate on large timescales}
\label{nonmon}
Variations of the pulsar rotational frequency may be complicated -- periodic, 
quasi-periodic, or completely stochastic. Generally, it may be described as a 
superposition 
\begin{equation}
\nu(t)=\nu_{ev}(t) + \delta\nu(t),
\label{eq1}
\end{equation}
where $\nu_{ev}(t)$ describes the secular evolution of pulsar parameters and 
$\delta\nu(t)$ corresponds to irregular variations. Similar expressions 
describe the evolution of $\dot \nu$ and $\ddot \nu$ after a differentiation.
The $\delta\ddot\nu(t)$ satisfies the obvious condition of zero mean value
$<\delta\ddot\nu(t)>_{t}\sim0$ over the timespans larger than characteristic
timescale of the variations. The amplitude of the observed variations of 
$\ddot\nu$ is related to the dispersion  of this process as 
$\sigma_{\delta\ddot\nu} = A_{\ddot\nu} = \sqrt{<(\delta\ddot\nu)^2>}$.

The second derivative values on the upper $\ddot\nu_{+}$ and lower 
$\ddot\nu_{-}$ branches in Fig. \ref{fig1} may be described as
$\ddot\nu_{\pm}(t) = \ddot\nu_{ev}(t) \pm A_{\ddot\nu}(t)$ for each pulsar. 
This equation describes some "average" pulsar, while the spread of points 
inside the branches reflects the variations of individual parameters over the 
pulsar ensemble and reaches 4 orders of magnitude.

The second derivative $\ddot \nu$ is the only parameter significantly 
influenced by the timing variations (see Section \ref{disc}). Thus one can 
assume that the measured values of $\nu$ and $\dot \nu$ may be considered to 
be evolutionary ones, $\nu_{ev}$ and $\dot \nu_{ev}$ (since $\delta \nu$ and 
$\delta \dot \nu$ are small).

\begin{figure}[t]
{\centering \resizebox*{1\columnwidth}{!}{\includegraphics[angle=270]
{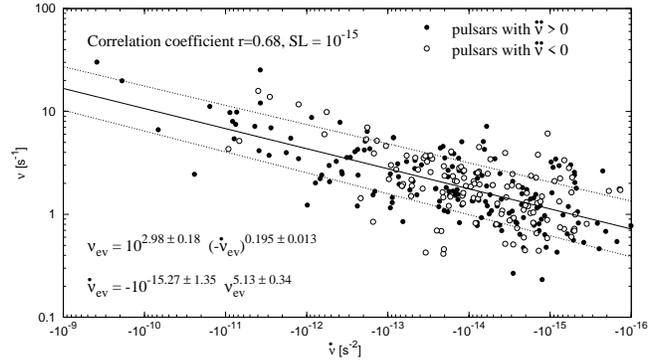}} \par}
\caption{The $\dot \nu - \nu$ diagram for the pulsars with the measured
second derivative. The filled symbols are objects with positive $\ddot\nu$,
the open ones -- with negative. The behavior of both subsets is the same.
The solid line represents the best fit, corresponding to the $n\approx5$
braking index, the dotted ones -- 1-$\sigma$ range.}
\label{fig2}
\end{figure}

Using relations described above, the secular behavior $\nu(t)$ (or, 
$\nu(\dot\nu)$) may be found by plotting the studied pulsar group onto the 
$\dot \nu - \nu$ diagram (Fig. \ref{fig2}). The objects with $\ddot \nu > 0$ 
and $\ddot \nu < 0$ are marked as filled and open circles, correspondingly. 
It is easily seen that the behavior of these two sub-groups is the same, 
which is in agreement with the smallness of the pulsar frequency variations 
in respect to the intrinsic scatter of $\nu(\dot \nu)$. However, a strong 
correlation between $\nu$ and $\dot\nu$ ($r\approx0.7)$ is seen, and
\begin{equation}
\dot\nu=-C\nu^n,
\label{eq2}
\end{equation}
where $C=10^{-15.26\pm1.38}$ and $n=5.13\pm0.34$. So, the secular evolution 
of the ``average'' pulsar is according to the ``standard'' spin-down law with 
$n\approx5$! This result is very interesting on its own, especially since 
the $\nu$ and $\dot \nu$ are always measured as independent values. The 
braking index $\approx 5$ may suggest the importance of multipole components 
of pulsar magnetic field, or the deviation of the angle between pulsar 
rotational and dipole axes from ${\pi}/{2}$ \citep{man77}. 

Note here, that the width of the fit on Fig. \ref{fig2} is quite small -- 
only about 1.5 orders of magnitude. If the spin-down is even approximately 
close to being described by the vacuum dipole model, then $C$ should scale as 
$(B_{0} \sin \alpha)^2$, where $B$ is the polar field and $\alpha$ is the 
magnetic inclination angle. But the range of $(B_{0}\sin\alpha)^2$ over the 
pulsar population is expected to be of many orders of magnitude. Therefore, 
Fig. \ref{fig2} once again shows that simple vacuum dipole model is not 
adequate to observations. The braking index $n \approx 5$ allows the 
interpretation of $C$ in terms of the quadrupole spin-down. It could be 
explained also in the frame of electric current mechanism of pulsar 
spin-down \citep{gur93}.

As has been stated above, the dependency between $\dot \nu$ and $\tau_{ch}$
is consistent with a power law with a slope of $-1.16 \pm 0.02$. This value 
is significantly different from $-1.0$ which would be the case when the 
$\nu - \dot \nu$ correlation is absent 
and roughly consists with a slope 
of 5.13 in the $\dot \nu(\nu)$ dependency (the values are different on a
$2.5\sigma$ level). But $n$ is strongly dependent on the $\dot\nu - \tau$ 
slope value, so $-1.16$ seems to be more or less consistent with the 
measured value of $n$.
                 
\begin{figure}[t]
{\centering \resizebox*{1\columnwidth}{!}{\includegraphics[angle=270]
{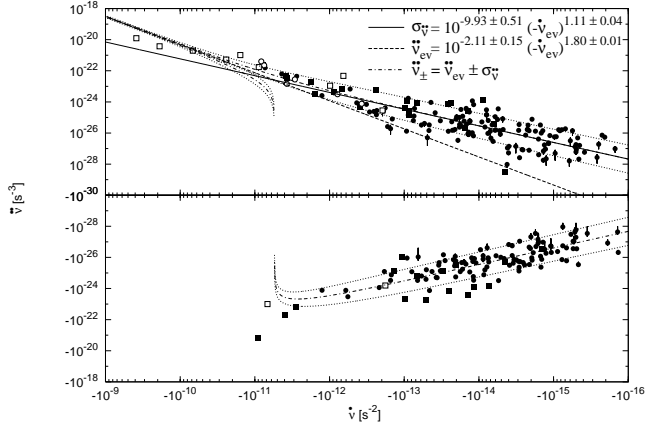}} \par}
\caption{The $\ddot \nu - \dot \nu$ diagram with the simple variations 
model. The solid line is the amplitude $A_{\ddot\nu}$ of the frequency 
second derivative variations, the dashed -- the secular term $\ddot\nu_{ev}$,
and the dot-dashed lines are the envelopes of the oscillations 
$\ddot\nu_{ev}\pm A_{\ddot\nu}$ with 1-$\sigma$ ranges (dotted lines). 
Each pulsar spends the majority of its life time at or very near the 
envelopes.}
\label{fig3}
\end{figure}

From (\ref{eq2}) we may easily determine the relation between $\ddot\nu_{ev}$ 
and $\dot\nu$ as $\ddot\nu_{ev} = nC^{\frac1n}(-\dot\nu)^{2-\frac1n}$, which 
is shown in Fig. \ref{fig3} as a thick dashed line. The same relation may be 
also estimated directly by using the asymmetry of the branches seen on Fig. 
\ref{fig1} as 
${\ddot\nu(\dot\nu) = \frac12\left(\ddot\nu_{+} + \ddot\nu_{-}\right)}$, 
where $\ddot \nu_{\pm}$ are defined in Fig. \ref{fig1}. Such estimation, 
while being very noisy, is positive in the 
$-10^{-11} \div -10^{-15}$ s$^{-2}$ range and agrees quantitatively with 
the previous one.

The amplitude of the $\ddot \nu$ oscillations, $A_{\ddot \nu}$, may be easily 
computed in a similar way, by using  $\ddot\nu_{+}$ and $\ddot\nu_{-}$, as 
$A_{\ddot \nu} = \frac12(\ddot \nu_{+} - \ddot\nu_{-})$.

The behaviour of pulsars according to the derived relations is shown in Fig. 
\ref{fig3}. This simple variations model describes the observed branches, 
both positive and negative, rather well. The absence of negative branch 
objects with $\dot\nu>-10^{-11}$ s$^{-2}$, i.e. with $\tau_{ch}>10^4$, we 
interpret as a prevalence of the second derivative's secular component over 
the varying one ($A_{\ddot\nu} < \ddot\nu_{ev}$) in this region. Older pulsars 
begin to change the sign of $\ddot\nu$ due to spin rate variations.

\section{Discussion and conclusions}
\label{disc}
In general, it is impossible to estimate the amplitudes of the frequency and 
its first derivative variations $A_{\nu}$ and $A_{\dot\nu}$ from the amplitude
of the second derivative only (the knowledge of its complete power density 
spectrum is needed).However, if the spectral density is relatively localized 
and some characteristic timescale $T$ of the variations exists, it is possible
to set some limits on it. A rough estimation is 
$A_{\nu}\sim A_{\dot\nu}T$, $A_{\dot\nu}\sim A_{\ddot\nu}T$ and 
$A_{\nu}\sim A_{\ddot\nu}T^2$. On a large timescale the variations can not 
lead to pulsar spin-up, so the variations of frequency first derivatives are 
much smaller than the secular ones, and 
${A_{\dot\nu}\sim A_{\ddot\nu}T \ll \dot\nu}$, so $T\ll\dot\nu/A_{\ddot\nu}$.
So, for the $-10^{-12} < \dot\nu < -10^{-15}$ s$^{-2}$ range and corresponding
values of $A_{\ddot\nu}$ from $10^{-23}$ s$^{-3}$ to $10^{-26}$ s$^{-3}$, the 
characteristic timescale $T_{up} \sim 10^{11}$ s. Also, this characteristic 
timescale is obviously larger than the timespan of observations, so 
$50 < T < 3\cdot10^3$ years. Assuming the constancy of $T$ during the pulsar 
evolution and therefore the change of $A_{\nu}$ with time, we get
$A_{\nu} \sim 10^{-3} \div 10^{-7}$ Hz. For such a model the pulsar 
frequency varies with the characteristic time of several hundred years and 
the amplitude from $10^{-3}$ Hz for young objects to $10^{-7}$ Hz for older 
ones.

The physical reasons of the discussed non-monotonic variations of the pulsar
spin-down rate may be similar to the ones of the timing noise on a short 
timescale. Several processes had been proposed for their explanation 
\citep{cor81} -- from the collective effects in the neutron star superfluid 
core to the electric current fluctuations in the pulsar magnetosphere. 
Whether these processes are able to produce long timescale variations is 
yet to be analyzed. On a short timescale, the pulsars show different timing 
behaviour. But on the long timescale their behavior seems to be alike.

The argument in favor of the similarity between the discussed variations and
the timing noise is the coincidence of the timing noise $\ddot\nu$ amplitude
extrapolated according to its power spectrum slope \citep{bay99} to the time
scale of hundreds of years, with the $A_{\ddot\nu}$ derived from our analysis 
for the same $\dot\nu$, i.e. the same ages (see Fig. 3).

At the same time, there are several low noise pulsars with large or negative
$\ddot\nu$. For 19 of 45 pulsars studied in \citep{d'a95} the timing noise 
is nearly absent (RMS $< 1\cdot10^{-3} P$). Six of them have anomalous 
$\ddot\nu$ measured in \citep{hob04}, which are well consistent with the 
$|\ddot\nu|$ - $\dot \nu$ correlation \citep{cor85, arz94} and have a wide 
range of $\dot \nu$. This shows a possible difference between timing noise 
and the long timescale variations described above.

In any case, the principal point is that all the pulsars evolve with
long-term variations, and the timescale of such variations significantly exceeds 
several tens of years. That explains the anomalous values of the observed 
$\ddot\nu$ and braking indices and gives reasonable values of the underlying 
secular spin-down parameters. 


\end{document}